\documentstyle[aps,prl,epsfig,manuscript,floats]{revtex}    
\newcommand{\pe}{$\pi^{0} \rightarrow e^{+} e^{-} e^{+} e^{-} \;$}
\newcommand{\ke}{$K_{L} \rightarrow e^{+} e^{-} e^{+} e^{-} \;$}
\newcommand{\kpe}{$K_{L} \rightarrow \pi^{0} \pi^{0} \pi^{0}_{DD} \;$}
\newcommand{\ksd}{$K_{L} \rightarrow \pi^{0} \pi^{0}_{D} \pi^{0}_{D} \;$}

\newcommand{\kete}{$K_{L} \rightarrow \pi^{\pm} e^{\mp} \nu \: e^{+} e^{-} \;$ }

\begin{document}
\draft          
\title{Measurements of the rare decay \ke.}

\author{
\parindent=0.in
\parindent=0.in
A.~Alavi-Harati$^{12}$,
T.~Alexopoulos$^{12}$,
M.~Arenton$^{11}$,
K.~Arisaka$^2$,
S.~Averitte$^{10}$,
A.R.~Barker$^5$,
L.~Bellantoni$^7$,
A.~Bellavance$^9$,
J.~Belz$^{10}$,
R.~Ben-David$^{7}$,
D.R.~Bergman$^{10}$,
E.~Blucher$^4$, 
G.J.~Bock$^7$,
C.~Bown$^4$, 
S.~Bright$^4$,
E.~Cheu$^1$,
S.~Childress$^7$,
R.~Coleman$^7$,
M.D.~Corcoran$^9$,
G.~Corti$^{11}$, 
B.~Cox$^{11}$,
M.B.~Crisler$^7$,
A.R.~Erwin$^{12}$,
R.~Ford$^7$,
A.~Glazov$^4$,
A.~Golossanov$^{11}$,
G.~Graham$^{4}$, 
J.~Graham$^4$,
K.~Hagan$^{11}$,
E.~Halkiadakis$^{10,\dagger}$,
J.~Hamm$^1$,
K.~Hanagaki$^{8}$,  
S.~Hidaka$^8$,
Y.B.~Hsiung$^7$,
V.~Jejer$^{11}$,
D.A.~Jensen$^7$,
R.~Kessler$^4$,
H.G.E.~Kobrak$^{3}$,
J.~LaDue$^5$,
A.~Lath$^{10}$,
A.~Ledovskoy$^{11}$,
P.L.~McBride$^7$,
P.~Mikelsons$^5$,
E.~Monnier$^{4,*}$,
T.~Nakaya$^{7}$,
K.S.~Nelson$^{11}$,
H.~Nguyen$^7$,
V.~O'Dell$^7$, 
M.~Pang$^7$, 
R.~Pordes$^7$,
V.~Prasad$^4$, 
B.~Quinn$^{4}$,
E.J.~Ramberg$^7$, 
R.E.~Ray$^7$,
A.~Roodman$^{4}$, 
M.~Sadamoto$^8$, 
S.~Schnetzer$^{10}$,
K.~Senyo$^{8}$, 
P.~Shanahan$^7$,
P.S.~Shawhan$^{4}$,
J.~Shields$^{11}$,
W.~Slater$^2$,
N.~Solomey$^4$,
S.V.~Somalwar$^{10}$, 
R.L.~Stone$^{10}$, 
E.C.~Swallow$^{4,6}$,
S.A.~Taegar$^1$,
R.J.~Tesarek$^{10}$, 
G.B.~Thomson$^{10}$,
P.A.~Toale$^5$,
A.~Tripathi$^2$,
R.~Tschirhart$^7$,
S.E.~Turner$^2$ 
Y.W.~Wah$^4$,
J.~Wang$^1$,
H.B.~White$^7$, 
J.~Whitmore$^7$,
B.~Winstein$^4$, 
R.~Winston$^4$, 
T.~Yamanaka$^8$,
E.D.~Zimmerman$^{4}$ \\
\vspace*{0.1in}
\footnotesize
{\it
$^1$ University of Arizona, Tucson, Arizona 85721 \\
$^2$ University of California at Los Angeles, Los Angeles, California 90095 \\
$^{3}$ University of California at San Diego, La Jolla, California 92093 \\
$^4$ The Enrico Fermi Institute, The University of Chicago, 
Chicago, Illinois 60637 \\
$^5$ University of Colorado, Boulder, Colorado 80309 \\
$^6$ Elmhurst College, Elmhurst, Illinois 60126 \\
$^7$ Fermi National Accelerator Laboratory, Batavia, Illinois 60510 \\
$^8$ Osaka University, Toyonaka, Osaka 560-0043 Japan \\
$^9$ Rice University, Houston, Texas 77005 \\
$^{10}$ Rutgers University, Piscataway, New Jersey 08854 \\
$^{11}$ The Department of Physics and Institute of Nuclear and 
Particle Physics, University of Virginia, 
Charlottesville, Virginia 22901 \\
$^{12}$ University of Wisconsin, Madison, Wisconsin 53706 \\
$^{*}$ On leave from C.P.P. Marseille/C.N.R.S., France \\
$^{\dagger}$ To whom correspondence should be addressed. \\
}
\vspace*{0.1in}
\centerline{ \bf (The KTeV Collaboration)}
\vspace*{0.5in}
\centerline{ \bf Abstract}
\vspace*{0.2in}
\parbox{14cm}{ 
We observe 441 \ke candidate events with a background
of 4.2 events and measure B(\ke)$\; = (3.72 \pm 0.18(\mbox{stat}) \pm
0.23(\mbox{syst})) \times 10^{-8}$ in the KTeV/E799II experiment at
Fermilab.  Using the distribution of the angle between the planes of
the $e^{+} e^{-}$ pairs, we measure the CP parameters $\beta_{CP} =
-0.23 \pm 0.09(\mbox{stat}) \pm 0.02(\mbox{syst})$ and $\gamma_{CP} =
-0.09 \pm 0.09(\mbox{stat}) \pm 0.02(\mbox{syst})$.  We also present the
first detailed study of the $e^{+} e^{-}$ invariant mass spectrum in
this decay mode.  
 }
\vspace*{-0.4in}
\normalsize
}

\maketitle

\newpage
{ The rare decay \ke proceeds via a two-virtual-photon intermediate
state with internal photon conversions to $e^{+}e^{-}$ pairs.  This
permits a measurement of the the $K_{L}\gamma^{*}\gamma^{*}$ form
factor, which is necessary to understand the long distance
contributions to other rare $K_{L}$ decays, in particular $K_{L}
\rightarrow \mu^{+} \mu^{-}$ \cite{mumu}, whose short distance
processes are sensitive to the CKM matrix element $V_{td}$.
The QED prediction \cite{mt}, neglecting radiative corrections and a
form factor, for the ratio ${{\rm B}(K_{L} \rightarrow e^{+} e^{-}
e^{+} e^{-})} / {{\rm B}(K_{L} \rightarrow \gamma \gamma)}$, together
with the experimental measurement of ${\rm B}(K_{L} \rightarrow \gamma
\gamma)$ \cite{pdg}, gives a branching ratio of $(3.65 \pm 0.09)\times
10^{-8}$.  The most precise previous measurements with \ke were based
on 27 candidate events \cite{guping}. We report here a measurement of
the branching ratio of the rare decay $K_{L} \rightarrow e^{+} e^{-}
e^{+} e^{-}$ at the Fermilab experiment E799II.  We also present a
study of CP symmetry in this decay and the first analysis of a
$K_{L}\gamma^{*}\gamma^{*}$ form factor.

The data were collected in 1997 with the KTeV detector at Fermilab.  The
detector components are described in \cite{ppee,eh}.  During E799II running
conditions, 800 GeV protons struck a BeO target; collimators and
magnets defined two nearly parallel $K_{L}$ beams, which decayed in a
65 m long region held under vacuum.  The decay region was surrounded
by photon veto detectors designed to detect particles escaping the
fiducial volume of the detector.  A charged particle spectrometer,
consisting of four drift chambers and an analyzing magnet, was used to
determine the charge, momentum, and trajectories of particles.  A pure
CsI calorimeter with 3100 crystals was used for photon detection and
charged particle identification.

The four-track trigger used to collect this dataset is described in detail in
\cite{ppee}.  In the offline analysis, we require two positively charged
tracks and two negatively charged tracks from a common vertex.  At
least three tracks must strike the CsI; the fourth may strike the CsI
or pass through one of the CsI beamholes.  A track is identified as an
$e^{\pm}$ if its $E/P$, the ratio of the energy measured by the
calorimeter to the momentum measured by the charged spectrometer, is
between 0.9 and 1.1.  We require the energy of the clusters deposited
in the calorimeter to be greater than $2$ GeV.  We also place cuts on
the track and vertex quality and require the photon veto energies to
be no more than the level of typical accidental activity.

To select \ke events we require the decay vertex to be within the
fiducial region of $95-155 \; \mbox{m}$ from the target and the total
momentum of the kaon in the lab to be between $25-215 \;
\mbox{GeV}/c$.  Figure~\ref{mpt2} shows the distribution of the
${P_T}^{2}$, the square of the component of the total momentum of the
daughter particles ($e^{+} e^{-} e^{+} e^{-}$) transverse to the kaon
line of flight, versus the four body invariant mass for the data
sample and for a Monte Carlo simulation.  We define the signal region
by ${P_T}^{2}$ less than $300 \; {(\mbox{MeV/c})}^{2}$ and $|M_{e^{+}
e^{-} e^{+} e^{-}} - M_{K^{0}}|$ less than $30 \; \mbox{MeV}/{c^{2}}$,
with $90\%$ efficiency.  The liberal cut on $M_{e^{+} e^{-} e^{+}
e^{-}}$ was chosen to retain radiative events $K_{L} \rightarrow e^{+}
e^{-} e^{+} e^{-} (\gamma)$ with an undetected soft photon.

A source of background to \ke is $K_{L} \rightarrow e^{+} e^{-}
\gamma$ or $K_{L} \rightarrow \gamma \gamma$, with one or two photon
conversions in the detector material.  The photon conversion
probability in the material upstream of the first drift chamber in our
detector is ${(2.74 \pm 0.11)} \times 10^{-3}$
\cite{edz}.  Removing events that have a minimum track separation at
the most upstream drift chamber of less than $1~$mm reduces this
background to $3.7 \pm 0.3$ events, estimated from Monte Carlo.
Another source of background is
\kete where the pion misses the calorimeter because of the beamholes
and is incorrectly assumed to be an electron.  The kinematic limit for the
invariant mass with the pion misidentified as an electron is $478~
\mbox{MeV}/c^{2}$.  Since B($K_{L} \rightarrow \pi^{\pm} e^{\mp} \nu
\: e^{+} e^{-}$) is not measured, we estimate it from the branching
ratio of the parent decay $K_{L} \rightarrow \pi^{\pm} e^{\mp} \nu \:
\gamma$.  This background is seen in the lower mass region in
Fig.~\ref{mpt2} (top) and is estimated to be $0.5 \pm 0.5$ events.
After all selection criteria are applied, we observe a total of 441
\ke candidate events and estimate the total background to be 4.2
events, as described above.  The distribution of the invariant mass of
$e^{+} e^{-} e^{+} e^{-}$ for the data and the Monte Carlo is shown in
Fig.~\ref{m4e}.

The distribution of the $e^{+} e^{-}$ invariant masses in the \ke
decay reveals the internal structure of the $K_{L}
\gamma^{*}\gamma^{*}$ vertex.  A related decay, $K_{L} \rightarrow
e^{+} e^{-} \gamma$, probes the $K_{L} \gamma \gamma^{*}$ vertex; the
form factor for that decay has been parametrized by Bergstr\"{o}m,
Mass\'{o}, and Singer (BMS) \cite{bms,ohl},
\newpage
\begin{eqnarray}
f(x) = {1 \over {1 - x (m_{K}^{2}/m_{\rho}^{2})}} + {{C \alpha_{K^*}} \over {1 - x(m_{K}^{2}/m_{K^{*}}^{2})}} \times \nonumber \\
\left[ { {4 \over 3} - {1 \over {1 - x (m_{K}^{2}/m_{\rho}^{2})}} - {1 \over {9({1 - x (m_{K}^{2}/m_{\omega}^{2})})}} } \right. \nonumber \\
- \left. {2 \over {9({1 - x (m_{K}^{2}/m_{\phi}^{2})})}} \right], \nonumber
\end{eqnarray}
\noindent where $x \equiv {M_{ee}}^{2}/{M_{K}}^{2}$ and $M_{K}$, $M_{\rho}$, $M_{K^{*}}$, $M_{\omega}$ and $M_{\phi}$ are the invariant masses of the corresponding mesons. The parameter $\alpha_{K^*}$ describes the relative strength of an
intermediate pseudoscalar decay amplitude and a vector meson decay
amplitude.  The constant $C$ is determined using various coupling
constants and is equal to 2.3 \cite{bms,ohl}.  The BMS model predicts
$|\alpha_{K^*}| = 0.2-0.3$; note that $\alpha_{K^*}
\approx 0.3$, rather than zero, best approximates a pointlike form
factor ($f(x) = 1$) for this analysis.  Since there are two $e^{+}
e^{-}$ pairs in this decay, we use a factorized expression for the
form factor.

We fit for the BMS parameter $\alpha_{K^*}$ from our 441 event sample
using a likelihood function as in \cite{ll,eh} with one parameter.  The
likelihood function is based on the \ke matrix element which is a
function of all kinematic variables.  Our \ke Monte Carlo simulation
uses the matrix element calculation found in \cite{mt} and does not
include radiative corrections.  Since radiative corrections could
potentially have a large effect on the form factor measurement, in
reality we measure an effective parameter, $\alpha_{K^*}^{{\rm eff}}$
which takes into account both the radiative effects and the form
factor.  Using the likelihood function we obtain $\alpha_{K^*}^{{\rm
eff}} = -0.14 \pm 0.16(\mbox{stat}) \pm 0.15(\mbox{syst})$.
Figure~\ref{mee_ff} (a) shows the $x$ distribution for data and for
Monte Carlo using our measured form factor and using a pointlike form
factor.  The bin-by-bin ratios of data/Monte Carlo (Fig.~\ref{mee_ff}
(b) and (c)) further demonstrate the difference between our
measured value of $\alpha_{K^*}^{{\rm eff}}$ and $f(x) = 1$.  The
systematic uncertainty of 0.15 is due to the detector acceptance, and
is dominated by the uncertainty in reconstruction of the Z position of
the vertex and the calorimeter energy.  Our result is in agreement
with the recent measurement of $\alpha_{K^*}^{{\rm eff}}= -0.15 \pm
0.06(\mbox{stat}) \pm 0.02(\mbox{syst})$ for the decay $K_{L}
\rightarrow e^{+} e^{-} \gamma$ \cite{na48}, also ignoring radiative
corrections $^{1}$.  Note that the effect of radiative corrections on
$\alpha_{K^*}$ could well be different in the two modes.

We also study the form factor parametrization of D'Ambrosio, Isidori
and Portol\'{e}s (DIP) \cite{dip}.
This parametrization of the $K_{L} \gamma^{*}\gamma^{*}$ form factor
is a function of two real coefficients of linear and quadratic terms
($\alpha_{{\rm DIP}}$ and $\beta_{{\rm DIP}}$, respectively) and is relatively
model independent.  Using a factorization model \cite{dip} the
expectations for $\alpha_{{\rm DIP}}$ and $\beta_{{\rm DIP}}$ from a $\rho$ form
factor are $-1.2$ and $1.4$, respectively.  Due to the dominance of
low $M_{ee}$, \ke is much less sensitive to the second order term
($\beta_{{\rm DIP}}$) than the decay $K_{L}
\rightarrow e^{+} e^{-} \mu^{+} \mu^{-} \;$ for a similar sample size;
thus, only the first order term ($\alpha_{{\rm DIP}}$) is relevant in this
analysis.  We measure $\alpha_{{\rm DIP}}^{{\rm eff}} = -1.1 \pm
0.6(\mbox{stat})$.  This value is in agreement with our measured
value of $\alpha_{K^*}^{{\rm eff}}$ since the relation between the two is
simply $\alpha_{{\rm DIP}} \approx -1 + 2.8\alpha_{K^*}$ (obtained by a
Taylor expansion of the BMS form factor).

A first order Taylor expansion of the \ke form factors yields
a generic linear form factor $(1 + \alpha_{{\rm Taylor}}(x_{1} + x_{2}))$,
with $\alpha_{{\rm Taylor}} \approx 0.42 - 1.2\alpha_{K^*} \approx
-\alpha_{{\rm DIP}}/2.4$.  We measure an effective parameter
$\alpha_{{\rm Taylor}}^{{\rm eff}} = 0.5 \pm 0.3(\mbox{stat})$.

We measure the branching ratio of \ke by normalizing to a sample of
49 809 \ksd decays ($\pi^{0}_{D}$ refers to the Dalitz decay $\pi^{0}
\rightarrow e^{+} e^{-} \gamma$).  These events were collected using
the same trigger as for \ke.  Offline selection criteria are
similar to those used in the \ke analysis.  We require four photons,
the eight-body ($e^{+} e^{-} e^{+} e^{-} \gamma \gamma \gamma \gamma$)
${P_T}^{2}$ to be less than $800 \; {(\mbox{MeV/c})}^{2}$ and
$|M_{e^{+} e^{-} e^{+} e^{-} \gamma \gamma \gamma \gamma } -
M_{K^{0}}|$ to be less than $20 \; \mbox{MeV}/{c^{2}}$.  The minimum
track separation at the most upstream drift chamber is required to be
greater than $1~ \mbox{mm}$ to reduce conversion background, such as
$K_{L} \rightarrow \pi^{0} \pi^{0} \pi^{0}_{D}$ with a photon from one
of the non-Dalitz $\pi^{0}$'s converting to $e^{+} e^{-}$.  The
background to \ksd from photon conversions in detector material is
estimated to be $844 \pm 66$ events.  Another background is \kpe
($\pi^{0}_{DD}$ refers to the double Dalitz decay \pe) since both have
the same eight-particle final state.  There are 24 possible
combinations of $e^{\pm}$'s and $\gamma$'s that yield \ksd and 3
combinations that yield \kpe.  We separate the two modes based on the
best $\chi^{2}$ for the corresponding $\pi^{0}$ decay hypothesis.  In
addition, we require $M_{\gamma \gamma}$ to be between $127.5-142.5 \;
\mbox{MeV}/{c^{2}}$, $M_{e^{+} e^{-} \gamma}$ between $127.5-142.5 \;
\mbox{MeV}/{c^{2}}$ and we exclude the region of $M_{e^{+} e^{-} e^{+}
e^{-}}$ between $127.5-142.5 \; \mbox{MeV}/{c^{2}}$.  The number of
\kpe background events is estimated to be $21 \pm 3$ events.  With
these selection criteria, we estimate a total of 865 background events
for the normalization mode \ksd.

The \ke Monte Carlo, used to estimate detector acceptance,
incorporates the BMS form factor with $\alpha_{K^*}^{{\rm eff}} = -0.14$,
our measured value.  For the normalization mode \ksd, a QED
calculation of radiative corrections exists in literature
\cite{eegrad} and is included in our simulation.  We measure the
branching ratio of \ke to be $(3.72 \pm 0.18(\mbox{stat}) \pm
0.10(\mbox{syst}) \pm 0.20(\mbox{norm})) \times 10^{-8}$ (``norm''
refers to the measurement uncertainty in the normalization branching
ratio \cite{pdg}), consistent with the theoretical prediction
$3.65\times 10^{-8}$ \cite{mt} and the most precise published experimental
result $(3.96 \pm 0.78(\mbox{stat}) \pm 0.32(\mbox{syst}))
\times 10^{-8}$ \cite{guping}.  Varying $\alpha_{K^*}^{{\rm eff}}$ by our measured
uncertainty of $\pm 0.23$ yields a systematic uncertainty in the
\ke branching ratio of $0.9\%$.  The addition of final state
bremsstrahlung in the Monte Carlo simulation does not affect this
result significantly. We assign a $1.1 \%$ error due to limited Monte
Carlo statistics and a $2.2 \%$ error due to uncertainties in detector
acceptance, which is dominated by the uncertainty in the track
reconstruction efficiency.  The $5.5\%$ uncertainty in the $K_{L}
\rightarrow \pi^{0} \pi^{0}_{D} \pi^{0}_{D}$ branching ratio
\cite{pdg} results in an additional systematic error.  The combined
systematic error is $6.1\%$.  Our measurement of the ratio ${{{\rm B}(K_{L}
\rightarrow e^{+} e^{-} e^{+} e^{-})} / {{\rm B}(\pi^{0} \rightarrow
e^{+} e^{-} \gamma)^{2}}} = (2.59 \pm 0.12(\mbox{stat}) \pm
0.07(\mbox{syst})) \times 10^{-4}$ is unaffected by the large
uncertainty in the $\pi^{0}_{D}$ branching ratio measurement.

We search for CP violation in \ke by studying the angle $\phi$ between
the planes of the two $e^{+} e^{-}$ pairs in the kaon rest frame.  We
fit the $\phi$ distribution to \cite{guping}
\begin{eqnarray}
{{d\Gamma(K_{L} \rightarrow e^{+} e^{-} e^{+} e^{-})}\over{d\phi}} &\propto&
1 + \beta_{CP} \cos (2\phi) + \gamma_{CP} \sin (2\phi), \nonumber \\
\beta_{CP} = {{{1-|\epsilon r|^{2}} \over {1+|\epsilon r|^{2}}}B }&,& 
\gamma_{CP} = {{2 \; {\rm Re}(\epsilon r)} \over {1+|\epsilon r|^{2}}}C, \label{eq:bg}
\end{eqnarray}

\noindent where $\epsilon$ is the CP violating mixing parameter and $r$ (which is approximately unity) is the ratio of the amplitudes of $K_{1}$ and
$K_{2}$ decaying to $e^{+} e^{-} e^{+} e^{-}$. Ignoring CP violation,
i.e. neglecting $\epsilon$, Eq.~\ref{eq:bg} reduces to the
Kroll-Wada formula \cite{kw}, where the constant $B$ is $-(+)0.20$ for
an odd(even) CP eigenstate.  With the introduction of CP violation the
coefficient $\beta_{CP}$ of the $\cos (2\phi)$ term differs from $B$ and
the $\sin (2\phi)$ term, $\gamma_{CP}$, no longer vanishes.
Unfortunately, there are no predictions for the constant $C$, and
therefore for $\gamma_{CP}$.

We measure the CP parameters $\beta_{CP}$ and $\gamma_{CP}$ using the
distribution of the angle $\phi$.  For the purposes of this
measurement, we choose the $e^{+} e^{-}$ pairing that minimizes the
product of the two $e^{+} e^{-}$ invariant masses.  In reality, both
pairings contribute to the matrix element since there are identical
particles in the final state; the pairing we choose is the dominant
contribution.  To optimize simultaneously the detector resolution of
$\phi$ and the statistical error, we require the invariant masses of
the $e^{+} e^{-}$ pairs, $M_{ee1}$ and $M_{ee2}$, to be greater than
$8\; \mbox{MeV}/{c^{2}}$, retaining 264 events.  Our expectation for $B$
becomes $-0.25$ as a result of discarding low $M_{ee}$ events.  We
measure $\beta_{CP} = -0.23 \pm 0.09(\mbox{stat}) \pm
0.02(\mbox{syst}) $ and $\gamma_{CP} = -0.09 \pm 0.09(\mbox{stat})
\pm 0.02(\mbox{syst})$ after correcting the $\phi$ distribution for
detector acceptance (Fig.~\ref{betgam}).  We estimate a systematic
error of 0.014 and 0.021 in $\beta_{CP}$ and $\gamma_{CP}$,
respectively, due to acceptance uncertainties and an error of 0.008 in
both $\beta_{CP}$ and $\gamma_{CP}$ due to detector resolution.  We
also set a $90\%$ confidence level (CL) limit of $|\gamma_{CP}| <
0.21$.  These results are consistent with the hypothesis that the
decay proceeds predominantly through the $CP=-1$ ($K_{2}$) state.
These parameters are insensitive to the effective form factor and to
the inclusion of final state bremsstrahlung in the Monte Carlo.

In summary, we observe 441 \ke candidate events with an estimated
background of 4.2 events.  We measure B(\ke)$ = (3.72 \pm
0.18(\mbox{stat}) \pm 0.23(\mbox{syst})) \times 10^{-8}$.  We also
study the form factor for the first time in \ke.  Our measurements of
$\alpha_{K^*}^{{\rm eff}} ({\rm BMS}) = -0.14 \pm 0.16(\mbox{stat})
\pm 0.15(\mbox{syst})$, $\alpha_{{\rm DIP}}^{{\rm eff}} = -1.1 \pm
0.6(\mbox{stat})$ and $\alpha_{{\rm Taylor}}^{{\rm eff}} = 0.5 \pm
0.3(\mbox{stat})$ describe the combined effects of a form factor and
of radiative effects.  We measure the CP parameters $\beta_{CP} =
-0.23 \pm 0.09(\mbox{stat}) \pm 0.02(\mbox{syst})$ and $\gamma_{CP} =
-0.09 \pm 0.09(\mbox{stat}) \pm 0.02(\mbox{syst})$ ($M_{ee} > 8 \;
\mbox{MeV}/c^{2}$), thus limiting $|\gamma_{CP}|$ to less than $0.21$
at $90\%$ CL.

\newpage
We gratefully acknowledge the support and effort of the Fermilab staff
and the technical staffs of the participating institutions for their
vital contributions.  This work was supported in part by the U.S.
Department of Energy, The National Science Foundation and The Ministry
of Education and Science of Japan.  

}

{
\begin{figure}[p!]
\begin{center}
   \epsfig{file=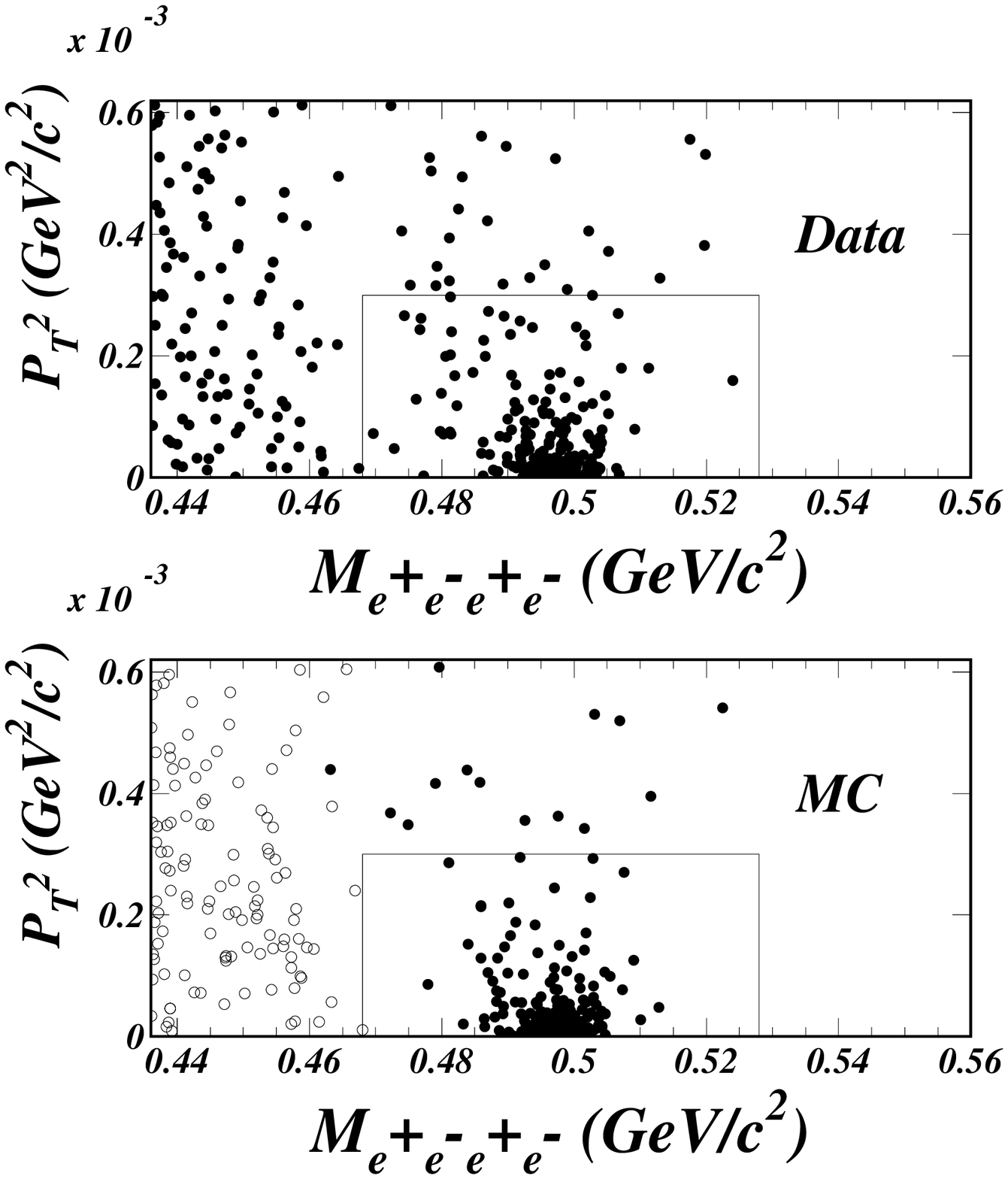,width=6.in,height=6.in}
\end{center} 
\caption{Top: The distribution of ${P_T}^{2}$ vs. $M_{e^{+} e^{-}
e^{+} e^{-}}$ for candidate \ke events after all cuts except the ${P_T}^{2}$
and $M_{e^{+} e^{-} e^{+} e^{-}}$ cuts.  There are 441 events in the signal
region defined by the box.  Bottom: The distribution of ${P_T}^{2}$
vs. $M_{e^{+} e^{-} e^{+} e^{-}}$ for reconstructed Monte Carlo
simulated events, scaled to the data statistics.  The filled circles
represent the signal Monte Carlo and the open circles represent the
\kete Monte Carlo.  The box defines the signal region with an
efficiency of $90\%$.}
\label{mpt2} 
\end{figure}
\begin{figure}[p!]
\begin{center}
   \epsfig{file=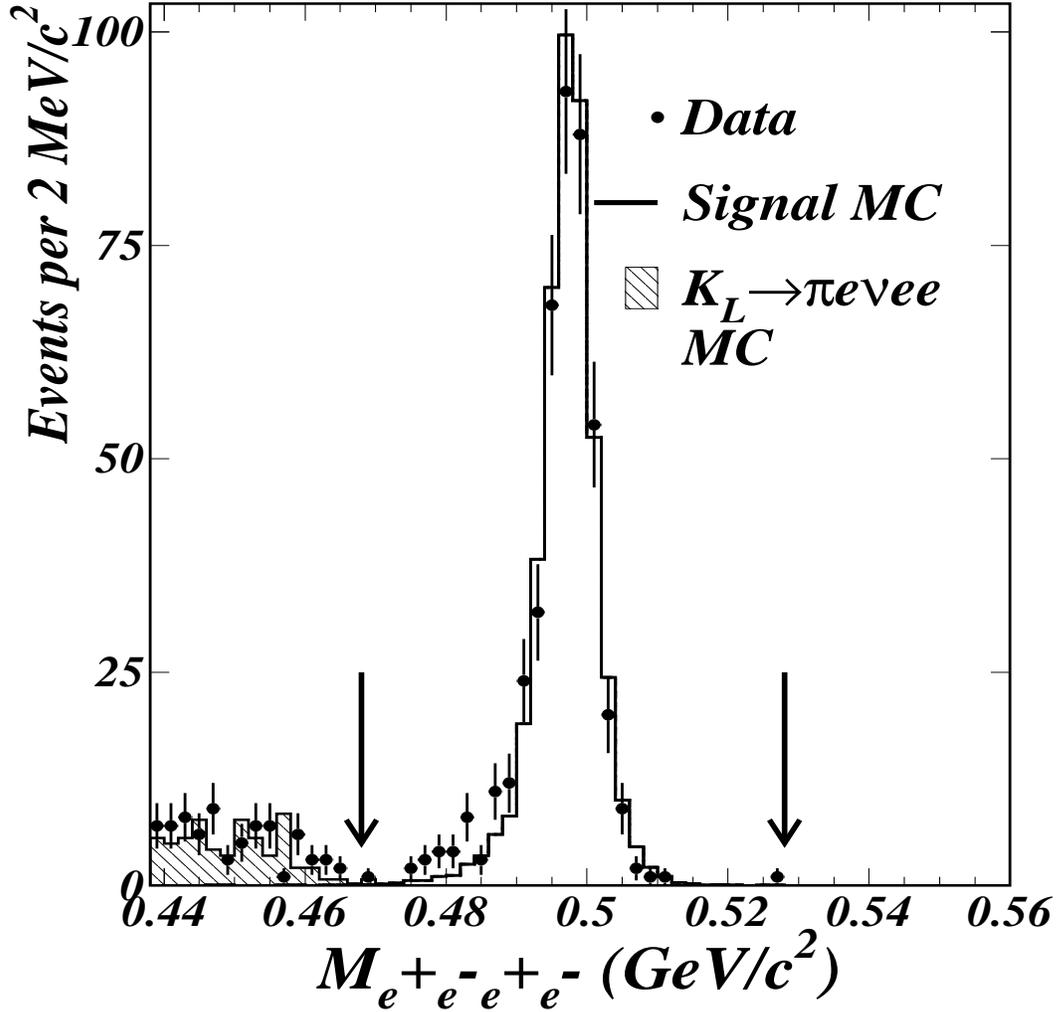,width=6.in,height=6.in}
\end{center} 
\caption{The $M_{e^{+} e^{-} e^{+} e^{-}}$ distribution after all cuts
except the invariant mass cut.  The dots represent the data and the
histogram represents the Monte Carlo simulation.  The arrows indicate
an intentionally wide mass window chosen to retain the low-side
radiative tail visible in the data.  The decay \kete is seen in the
lower mass region.}
\label{m4e} 
\end{figure}
\begin{figure}[p!]
\begin{center}
   \epsfig{file=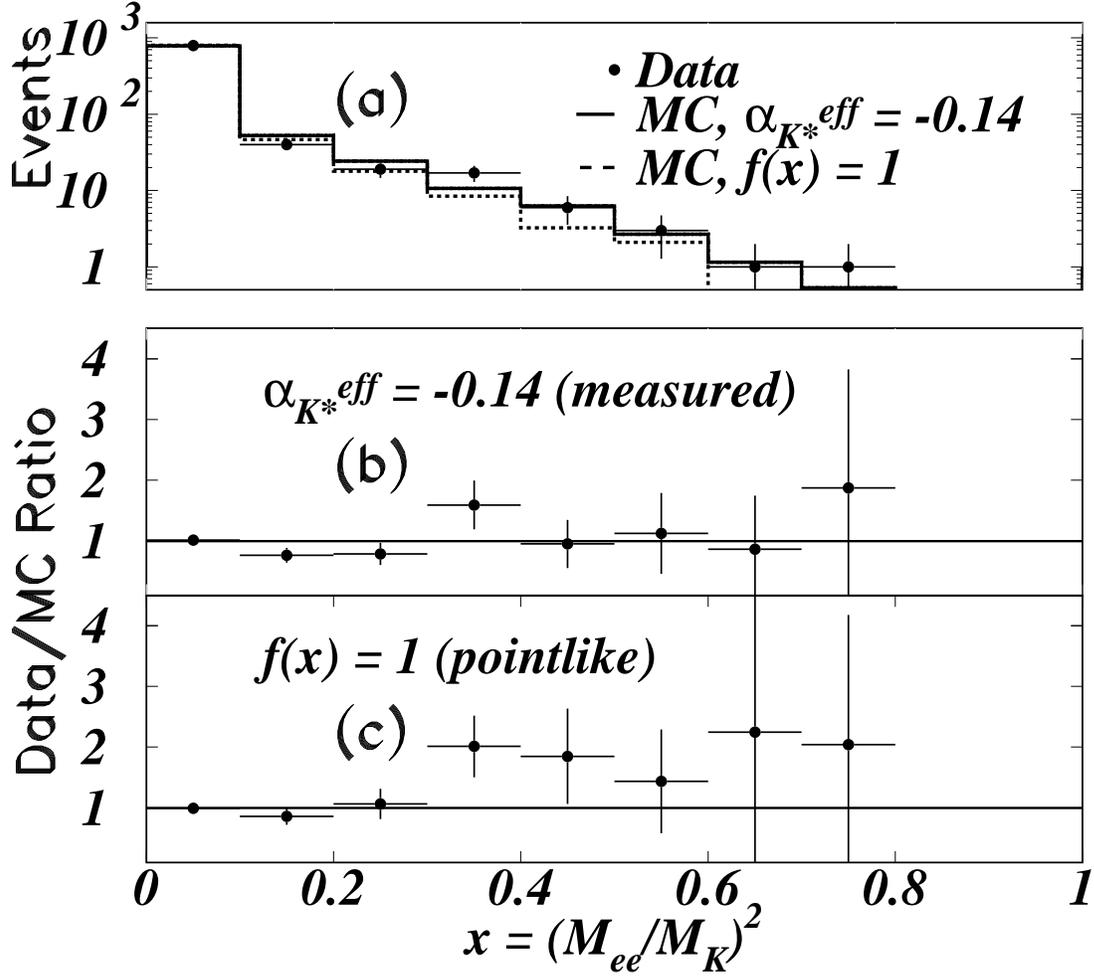,width=6.in,height=6.in}
\end{center}
\caption{(a) The $x$ distribution for data and for Monte Carlo using
our measured value of $\alpha_{K^*}^{{\rm eff}}$ and using $f(x) = 1$.
Since there are two $e^{+} e^{-}$ pairs in this decay, this is a
double entry plot.  The data/Monte Carlo ratio is shown for our
measured value of $\alpha_{K^*}^{{\rm eff}}$ (b) and for a pointlike
form factor (c).  We measure $\alpha_{K^*}^{{\rm eff}} = -0.14 \pm
0.16({\rm stat}) \pm 0.15({\rm syst})$.  Recall that $\alpha_{K^*}
\approx 0.3$ best approximates a pointlike form factor.}
\label{mee_ff}
\end{figure}
\begin{figure}[p!]
\begin{center}
   \epsfig{file=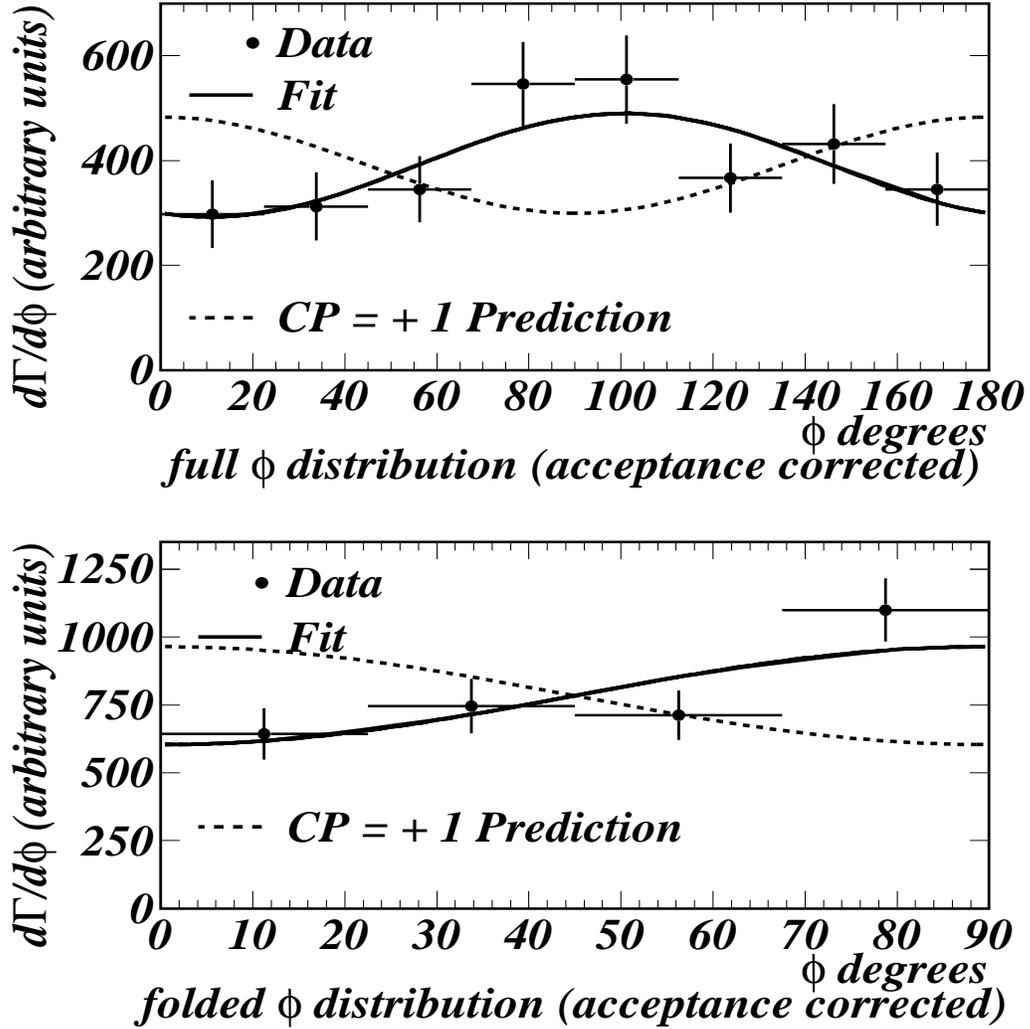,width=6.in,height=6.in}
\end{center} 
\caption{The distribution of the angle $\phi$ between the planes of
the two $e^{+} e^{-}$ pairs.  In the bottom figure, the $\phi$
distribution from $90^{\circ}-180^{\circ}$ is folded into
$0^{\circ}-90^{\circ}$.  We fit these distributions to Eq.~\ref{eq:bg}
and measure $\beta_{CP} = -0.23 \pm 0.09({\rm stat}) \pm 0.02({\rm
syst})$ and $\gamma_{CP} = -0.09 \pm 0.09({\rm stat}) \pm 0.02({\rm
syst})$.  The dashed line shows the $CP=+1$($K_{1}$) prediction and
further confirms that the decay proceeds predominantly through the
$CP=-1$ ($K_{2}$) state.}
\label{betgam} 
\end{figure}
}


\begin{thebibliography}{99}
\bibitem[1]{foot}  Including radiative corrections to $K_{L} \rightarrow e^{+}
e^{-} \gamma$, $\alpha_{K^*} = -0.36 \pm 0.06(\mbox{stat}) \pm
0.02(\mbox{syst})$ \cite{na48}.  The world average value for this
decay mode is $\alpha_{K^*} = -0.33 \pm 0.05$ \cite{pdg}.
\end{thebibliography}
\end{document}